\documentclass[journal]{IEEEtran}
\usepackage{multirow,epsfig,fbox,amsfonts,amsmath,multicol,enumitem,algorithm,algorithmic,pifont,graphicx,bm}
\usepackage{booktabs}
\usepackage{bm,enumitem}
\graphicspath{{figs/}}

\newcommand{\jl}[1]{\textcolor{black}{#1}}

\usepackage{subfigure}
\usepackage[numbers,sort&compress,comma]{natbib}
\usepackage{hyperref}
\hypersetup{
	colorlinks=true,
	linkcolor=blue
}

\graphicspath{{Figures/}}

\begin{document}
\title{Diffusion Enhancement for Cloud Removal in Ultra-Resolution Remote Sensing Imagery}
\author{
    Jialu~Sui,
    Yiyang~Ma,
    Wenhan~Yang,~\IEEEmembership{Member,~IEEE},
    Xiaokang~Zhang,~\IEEEmembership{Member,~IEEE},\\
    Man-On~Pun,~\IEEEmembership{Senior~Member,~IEEE},
    and 
    Jiaying~Liu,~\IEEEmembership{Senior~Member,~IEEE}

	\thanks{This work was supported in part by the Guangdong Provincial Key Laboratory of Future Networks of Intelligence under Grant 2022B1212010001, and the National Natural Science Foundation of China under Grant 42371374. \textit{(Corresponding authors: Man-On Pun)}}
	\thanks{Jialu Sui and Man-On Pun are with the School of Science and Engineering, The Chinese University of Hong Kong, Shenzhen, Shenzhen 518172, China (e-mail: jialusui@link.cuhk.edu.cn; SimonPun@cuhk.edu.cn).}
    \thanks{Yiyang Ma and Jiaying Liu are with the Wangxuan Institute of Computer Technology, Peking University, Beijing 100871, China (e-mail: myy12769@pku.edu.cn; liujiaying@pku.edu.cn).}
    \thanks{Wenhan Yang is with the Peng Cheng laboratory, Shenzhen, China (e-mail: yangwh@pcl.ac.cn).}
    \thanks{Xiaokang Zhang is with the School of Information Science and Engineering, Wuhan University of Science and Technology, Wuhan 430081, China (e-mail: natezhangxk@gmail.com).}
    }
\maketitle
  
\begin{abstract}
The presence of cloud layers severely compromises the quality and effectiveness of optical remote sensing (RS) images. However, existing deep-learning (DL)-based Cloud Removal (CR) techniques encounter difficulties in accurately reconstructing the original visual authenticity and detailed semantic content of the images. To tackle this challenge, this work proposes to encompass enhancements at the data and methodology fronts. On the data side, an ultra-resolution benchmark named CUHK Cloud Removal (CUHK-CR) of $0.5$~m spatial resolution is established. This benchmark incorporates rich detailed textures and diverse cloud coverage, serving as a robust foundation for designing and assessing CR models. From the methodology perspective, a novel diffusion-based framework for CR called Diffusion Enhancement (DE) is proposed to perform progressive texture detail recovery, which mitigates the training difficulty with improved inference accuracy. Additionally, a Weight Allocation (WA) network is developed to dynamically adjust the weights for feature fusion, thereby further improving performance, particularly in the context of ultra-resolution image generation. Furthermore, a coarse-to-fine training strategy is applied to effectively expedite training convergence while reducing the computational complexity required to handle ultra-resolution images. Extensive experiments on the newly established CUHK-CR and existing datasets such as RICE confirm that the proposed DE framework outperforms existing DL-based methods in terms of both perceptual quality and signal fidelity.
\end{abstract}

\begin{IEEEkeywords}
Denoising diffusion probabilistic model, Cloud removal, Remote sensing images
\end{IEEEkeywords}

\IEEEpeerreviewmaketitle

\section{Introduction}\label{sec:intro}

Remote sensing (RS) images play a crucial role in a variety of applications, including change detection \cite{changedetection}, semantic segmentation \cite{segmentation}, and object detection \cite{detection}. However, the imaging capabilities of satellite sensors, characterized by their ultra-long-range nature, make them susceptible to degradation, resulting in quality distortions in the captured images. One significant factor contributing to such degradation is the presence of cloud cover. Clouds significantly reduce visibility and saturation in the images, undermining the effectiveness of RS images, especially in the optical domain. This cloud-induced degradation hampers the clarity and detail of the images, impacting their practical utility. Consequently, there is a pressing need for the development of restoration methods aimed at enhancing land surface information obscured by cloud layers, thereby improving the effectiveness of remote sensing images.

Traditional methods for Cloud Removal (CR) can be broadly categorized into two main groups, namely multi-spectral and multi-temporal techniques. More specifically, multi-spectral methods \cite{spectral1, spectral2, spectral3, spectral4} primarily rely on variations in wavelength-dependent absorption and reflection to recover obscured landscapes caused by haze and thin cirrus clouds. However, in scenarios involving thick and filmy clouds that entirely obstruct optical signals, the efficacy of multi-spectral methods may be compromised due to the absence of supplementary information. In contrast, multi-temporal methods \cite{temporal1,temporal2} integrate clear sky conditions from reference images captured at different time instances. While the results derived from the multi-temporal methods are more reliable in general as they stem from actual cloud-free observations, the rapid changes in the landscape significantly impact the accuracy of the reconstructed images.

In recent years, deep-learning (DL)-based methods have gained significant popularity for their extraordinary ability to generate high-quality, cloud-free results. These approaches within the realm of DL can be further categorized into CNN-based models \cite{cvae}, Generative Adversarial Network (GAN)-based models \cite{gan,cyclegan}, and diffusion-based models \cite{crddpm}. More specifically, CNN-based models operate by inputting cloudy images into a network and updating parameters based on loss functions calculated from the output and the corresponding cloud-free image. Along the same direction, \citet{DSen2} introduced a deep residual neural network designed to reconstruct an optical representation of the underlying land surface structure. Notably, SAR imagery was incorporated into the CR process to offer additional information on surface characteristics beneath clouds. Additionally, \citet{cloudmat} utilized a two-step convolution network to extract transparency information from clouds and determine their positions. However, the feature representation capability of CNN-based models is constrained, limiting their ability to generate cloud-free images with superior perceptual quality.

To address this limitation, GAN-based models employ unique training strategies that incorporate two key components, namely the generator and the discriminator. The generator creates cloud-free images, while the discriminator evaluates whether the generated images meet desired quality standards, providing gradients for updating the generator's parameters through an additional GAN loss function. For instance, Cloud-GAN \cite{cyclegan} preserved color composition and texture by learning a bidirectional mapping of feature representation between cloudy images and their corresponding cloud-free counterparts in a cyclic structure. Nevertheless, GAN-based models grapple with persistent challenges, including model collapse, unstable training dynamics, and vanishing gradients, which adversely impacts their overall performance in diverse applications.

Recently, a novel branch of generative models, known as diffusion models \cite{DDPM}, has been introduced to computer vision tasks. These models have demonstrated remarkable performance in generating detailed textures across various low-level tasks, including super-resolution \cite{sui2023gcrdn,sui2023dtrn,ma2023solving}, deblurring \cite{deblurr,deblurr2}, and inpainting \cite{impaint}.  Optimal integration of the gradual learning and refinement features of diffusion models into the generation process is expected to pave the way for more advanced and effective approaches in CR. However, it is noteworthy that the outcomes obtained from pure diffusion models for CR are often inaccurate with undesirable fake textures. Consequently, the current applications of diffusion models in CR primarily focus on feature extraction \cite{crddpm}, limiting their inherent capabilities for gradual learning and refinement in this context.

In this study, based on the diffusion architecture, we propose a novel network named Diffusion Enhancement (DE) for CR, aiming to leverage the inherent strengths of the diffusion model to improve the quality of images. In sharp contrast to existing diffusion-based methods that only rely on progressive refinement for reconstructing fine-grained texture details, this work proposes to integrate the reference visual prior. In this way, the global visual information can be effectively integrated into the progressive diffusion process to mitigate the training difficulty, which results in improved inference accuracy. Besides, a weight allocation (WA) network is introduced to optimize the dynamic fusion of the reference visual prior and intermediate denoising images derived from the diffusion models. To expedite the diffusion model convergence, we further propose a coarse-to-fine training strategy. More specifically, the network is first trained on smaller patches before being fine-tuned using larger patches. Finally, taking advantage of recent satellite observations of high quality and resolution \cite{SR, resolution, xiaokang}, an ultra-resolution benchmark containing clear spatial texture information of the location and intrinsic features of the landscape is established for CR algorithm design and performance evaluation.

In summary, the main contributions of this work are summarized as follows:
\begin{enumerate}[wide, labelwidth=!, labelindent=0pt]
\item A novel network called Diffusion Enhancement is proposed in this work to restore land surface under cloud cover. The proposed DE network, which merges global visual information with progressive diffusion recovery, offers enhanced capability of capturing data distribution. As a result, it excels in predicting detailed information by utilizing reference visual prior during the inference process;

\item A weight allocation module is devised to compute adaptive weighting coefficients for the fusion of the reference visual prior and intermediate denoising images derived from the diffusion models. As a result, the reference visual prior refinement predominantly contributes to coarse-grained content reconstruction in the initial steps, while the diffusion model focuses its efforts on incorporating rich details in the subsequent stages. In addition, a coarse-to-fine training strategy is applied to stabilize the training while accelerating the convergence speed of DE;

\item Finally, an ultra-resolution benchmark called CUHK-CR is established to evaluate the CR methods against different types of cloud coverage. Our benchmark consists of $668$ images of thin clouds and $559$ images of thick clouds with multispectral information. To our best knowledge, our benchmark stands for the CR dataset of the highest spatial resolution, i.e. $0.5$~m, among all existing CR datasets. The data and code can be downloaded from GitHub\footnote{\href{https://github.com/littlebeen/Diffusion-Enhancement-for-CR}{https://github.com/littlebeen/Diffusion-Enhancement-for-CR}.}.
\end{enumerate}

The remainder of this paper is structured as follows: an overview on existing CR datasets and methods is first presented in Section~\ref{sec:related} before Section~\ref{sec:dataset} outlines in detail our dataset CUHK-CR. After that, Section~\ref{sec:method} introduces the proposed DE network whereas experimental findings and insights are deliberated in Section~\ref{sec:experiments}. Finally, concluding remarks are offered in Section~\ref{sec:conclusion}.

\begin{table*}[h]
	\centering
	\caption{Comparison between existing CR datasets and CUHK-CR.}
	\setlength{\tabcolsep}{1.3mm}{
		\begin{tabular}{ccccccc}
			\hline
                \centering\textbf{Dataset}&\textbf{Acquired Time gap}&\textbf{Image Size}&\textbf{Number}&\textbf{Resolution (m)}&\textbf{Spectrum}&\textbf{Source}\\
                \hline
                T-Cloud \cite{cvae}&16 days&256&2,939&30&3&Landsat 8\\
                RICE1 \cite{rice}&15 days&512&500&30&3&Google Earth\\
                RICE2 \cite{rice}&15 days&512&736&30&3&Landsat 8\\
                WHUS2-CR \cite{whus}&12 days&large&36&10&10&Sentinel-2A\\
                SEN12MS-CR \cite{sen12ms}&14 days&256&122,218&10&13&Sentinel-2\\
                WHU Cloud Dataset \cite{whuc}&6 months&512&859&30&3&Landsat 8\\
                \hline
                CUHK-CR1&17 days&512&668&0.5&4&Jilin-1\\
                CUHK-CR2&17 days&512&559&0.5&4&Jilin-1\\
		\hline
	\end{tabular}}
	\label{tab:datasets}
\end{table*}

\section{Related Work}\label{sec:related}
\subsection{Conventional end-to-end method for CR}
End-to-end Cloud Removal models are specifically designed to take a cloudy image as input and directly generate a cloud-free image during the inference process. These models excel in swiftly producing inference results, primarily focusing on discerning the differences between the cloudy image and its corresponding cloud-free counterpart. CVAE \cite{cvae} delves into the image degradation process using a probabilistic graphical model whereas SpAGAN \cite{spagan} emulates the human visual mechanism by employing a local-to-global spatial attention approach to detect and highlight cloud regions. Furthermore, AMGAN-CR \cite{amgan} removes clouds using an attentive residual network guided by an attention map. Despite their merits, the visual outcomes of these end-to-end models consistently replace clouds with neighboring colors, lacking the capability to predict the underlying texture obscured by clouds. This limitation adversely impacts the effectiveness of these CR methods, particularly in cases of dense cloud coverage.

\subsection{The diffusion architecture and prior guidance}
Recently, the diffusion model \cite{DDPM,latentddpm,dualdiff} has garnered significant attention. This model gradually generates the ultimate result, denoted as $\mathbf{x}_0$, from a latent variable $\mathbf{x}_T$, where $T$ represents the total number of diffusion steps in a parameterized Markov chain. The diffusion model comprises two key components, namely the forward process and the reverse process. More specifically, the forward process transforms the data distribution into a latent variable distribution through a step-by-step progression, leveraging the parameters of the Markov chain to transition from the initial data space to the latent space. Conversely, the reverse process aims to revert the latent variable distribution back to the original data distribution, recovering the initial data and providing a comprehensive understanding of the underlying data distribution.

In contrast to previously discussed end-to-end methods, the diffusion model \cite{crdiff2,crdiff1} offers a higher level of detailed information, beneficial for restoring the landscape under cloud coverage. However, the conventional diffusion model tends to generate unreliable fake textures. In the absence of effective solutions to this issue, current diffusion model-based methods like DDPM-CR \cite{crddpm} primarily employ the diffusion model as a feature extractor, which overlooks the potential to leverage the diffusion model's inherent strengths in gradual learning and refinement. Alternatively, some pioneering attempts \cite{prior1,prior2} have been made to incorporate prior guidance into the inference process. Aiming to fully exploit the diffusion model's potential for incremental learning and iterative refinement, the proposed DE network is crafted to improve the generation process by leveraging the reference visual prior.

\subsection{Datasets for CR}
Table~\ref{tab:datasets} lists several of the most representative existing image datasets for optical-based CR. As shown in Table~\ref{tab:datasets}, all the datasets share a common drawback, i.e. their low spatial resolution of around $10$ to $30$ meters. This limitation significantly compromises the level of spatial detail they can provide. Furthermore, despite the fact that the multispectral information is necessary for the satellite image analysis, datasets such as T-Cloud \cite{cvae} and RICE \cite{rice} only contain RGB bands. In addition, it is advantageous to minimize the ``Acquired time gap" as significant landscape changes can occur between the time instances of taking the cloudy image and its corresponding clear image. However, popular datasets like WHU Cloud Dataset \cite{whuc} possess large ``Acquired time gap", which can be an issue of concern in practice. Finally, all datasets listed in Table~\ref{tab:datasets} were generated with open-source satellites such as Landsat $8$ and Sentinel-$2$. It is highly desirable to have datasets from more satellites with different sensor characteristics for CR algorithm design and performance assessment. 

\section{Proposed CUHK-CR dataset}\label{sec:dataset}
\subsection{CUHK-CR}
Driven by the ever-increasing resolution of remote sensing imagery, we have established a new ultra-resolution benchmark named CUHK Cloud Removal (CUHK-CR). This benchmark is characterized by its ultra-high spatial resolution of $0.5$~m and $4$ multispectral bands with data acquisition confined to a period of $17$ days. Such an ultra-high spatial resolution benchmark can facilitate the training and evaluation of various CR methods specifically designed for ultra-resolution images. As a result, the benchmark can mitigate the gap between the low-resolution images during training and the high-resolution acquired in the real world, which is shown particularly critical for good CR performance in Section~\ref{sec:experiments}. Furthermore, the benchmark comprises two subsets, a thin cloud subset, namely CUHK-CR1 and a thick cloud subset, namely CUHK-CR2, facilitating training and evaluation on varying cloud coverage. More specifically, the thin cloud subset includes $668$ images, while the thick cloud subset $559$ images. These images were cropped into smaller segments for convenience, directly compatible with deep-learning models. Unless specified otherwise, a training-to-testing set ratio of $8:2$ is employed in the sequel, resulting in $534$ and $448$ images for training, and $134$ and $111$ images for testing in the thin and thick subsets, respectively. Finally, it is worth pointing out that our dataset is based on a new commercial satellite, Jilin-1, instead of those frequently utilized satellites like Landsat-8 and Sentinel. The distinct image contexts provided by the Jilin-1 satellite sensors contribute to the uniqueness of our dataset.

\subsection{Data Collection}

\begin{table}[t]
	\centering
	\caption{Jilin-1KF01B sensor bands.}
	\setlength{\tabcolsep}{1.3mm}{
		\begin{tabular}{ccc}
			\hline		
                \centering\textbf{Band Name}&\textbf{Spectral Coverage (nm)}&\textbf{Spatial Resolution (m)} \\
                \hline
                Panchromatic color&450-800&0.5\\
                Blue&450-510&2\\
                Green&510-580&2\\
                Red&630-690&2\\
                Near-infrared&770-895&2\\
		\hline
	\end{tabular}}
	\label{tab:color}
\end{table}

\begin{table*}[h]
	\centering
	\caption{The summary of the study sites of CUHK-CR.}
	\setlength{\tabcolsep}{1.3mm}{
		\begin{tabular}{cccccc}
			\hline
                \centering\textbf{Data}&\textbf{Location}&\textbf{Size (km$^2$)}&\textbf{Covers}&\textbf{Acquired Time}&\textbf{Temporal data Time}\\
                \hline
                I&Shenzhen Guangdong Province&35.72&City, Forest&2022/08/24& 2022/09/03\\
                III&Jinan Shandong Province&34.76&City, River&2022/08/24&2022/09/10\\
                III&Hangzhou Zhejiang Province&	25.62 &City, Lack&2023/01/31&2023/01/21\\
		\hline
	\end{tabular}}
	\label{tab:sample}
\end{table*}

Jilin-1 satellite constellation is the core project of Chang Guang Satellite Technology Co., Ltd. (CGSTL). The constellation is composed of $138$ high-performance optical RS satellites, covering high resolution, large width, video and multi-spectrum. Our dataset was collected by a satellite named Jilin-1KF01B equipped with a $0.5$~m resolution push broom camera. Launched in 2021, Jilin-1KF01B incorporates advanced technology to acquire more than 2 million km$^2$ of high-definition images every day with a width greater than $150$~km. As shown in Table \ref{tab:color}, the push broom camera covers four spectral bands, namely blue, green, red, and near-infrared, as well as a high-resolution panchromatic color band. Pen-sharpening using the complementary information from the multispectral and panchromatic images is applied to improve the spatial resolution of the spectral bands from $2$~m to $0.5$~m.

Table \ref{tab:sample} refers to the location, size, coverage, and acquired time of the cloudy images and their corresponding cloud-free images. The location of the satellite images is chosen from the north to the south of China while the gap in acquire time is limited to $17$ days.

\subsection{Data Analysis}
\begin{figure}[tbp]
  \centering
   \includegraphics[width=0.95\linewidth]{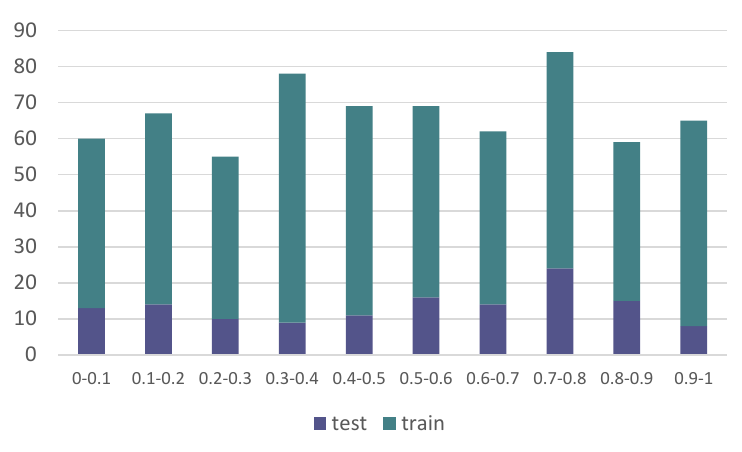}
   \caption{The \jl{distribution} of images on different CCP of CUHK-CR1 training and test dataset computed via the detector of  Cloud-Net \cite{clouddetection1}. The average probability of cloud coverage is 50.7\%.}
   \label{fig:PCC1}
\end{figure}
\begin{figure}[tbp]
  \centering
   \includegraphics[width=0.95\linewidth]{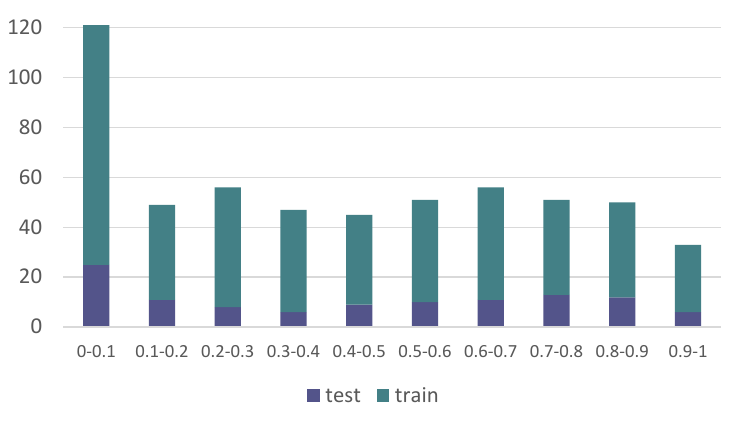}
   \caption{The \jl{distribution} of images on different CCP of CUHK-CR2 training and test dataset computed via the detector of Cloud-Net \cite{clouddetection1}. The average probability of cloud coverage is 42.5\%.}
   \label{fig:PCC2}
\end{figure}

To analyze the cloud coverage statistics in the CUHK-CR dataset, we calculate the widely used Cloud Coverage Probability (CCP) \cite{sen12ms} on two distinct sets. We visualize the distribution of image counts for different CCP values in Fig.~\ref{fig:PCC1} and Fig.~\ref{fig:PCC2}.

For each optical image, the Cloud-Net detector \cite{clouddetection1,38-cloud-1} is applied to produce binary masks with pixel-wise values of either $0$ or $1$ with $0$ and $1$ indicating the cloudy and cloud-free places, respectively. It is important to note that the detector fails to differentiate between thin and thick cloud layers. It simply detects the presence of cloud cover at the pixel level. Thin clouds typically extend over a broader area, whereas thick clouds occupy a smaller portion of the image, including richer reference information used for predicting background ground. We remove those images whose landscapes are totally obscured by the dense clouds through visual observation. As a result, the average CCP for the set with thin clouds is higher than that for the set with thick clouds. Notably, the images with CCP between $0$ to $0.1$ account for the largest proportion in the CUHK-CR2.

\begin{figure*}[tbp]
  \centering
   \includegraphics[width=\linewidth]{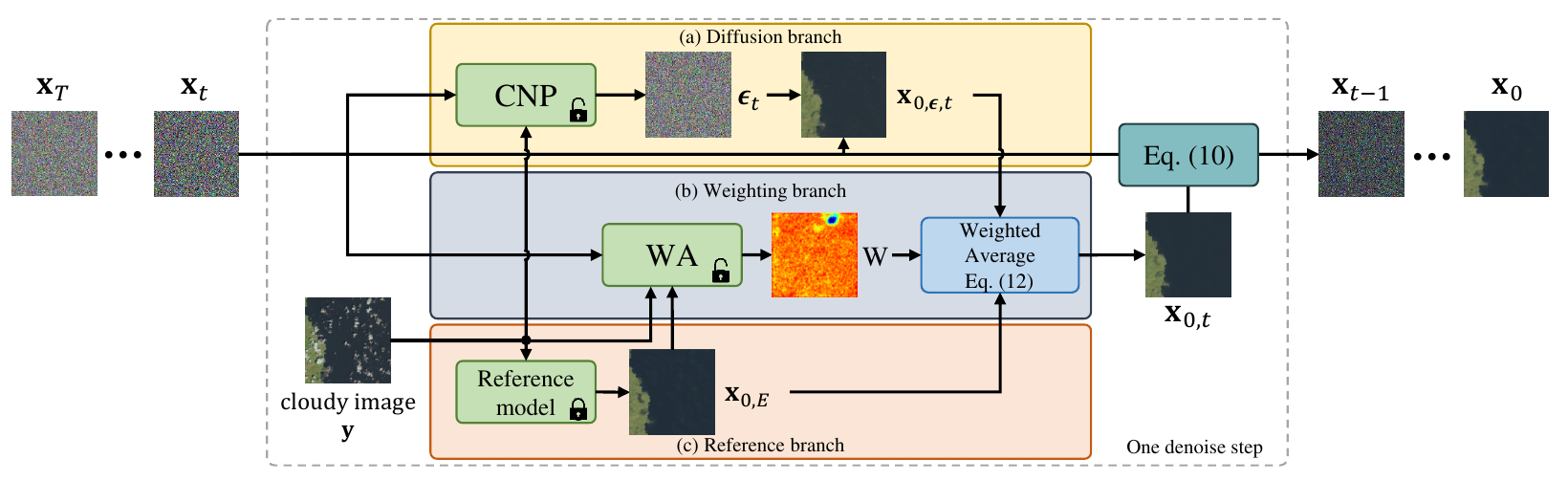}
   \caption{The architecture of our DE for CR. 
   	\textbf{Diffusion branch} in (a) performs the diffusion step that removes noise progressively, which is capable of restoring fine-grained textures.
   	\textbf{Weighting branch} in (b) performs the dynamic fusion of results from both the reference and diffusion branches with the result $\mathbf{x}_{0,t}$, capturing the merits of both excellent global estimations and fine details.
   	\textbf{Reference branch} in (c) generates a cloud-free image based on the cloudy image $\mathbf{y}$, offering substantial global context.
   	Ultimately, $\mathbf{x}_{0,t}$ and $\mathbf{x}_t$ are utilized in the generation of $\mathbf{x}_{t-1}$.
   }
   \label{fig:network}
\end{figure*}

\section{Diffusion Enhancement for Cloud Removal}\label{sec:method}
\subsection{Architecture}
Similar to the denoising diffusion probabilistic model \cite{DDPM}, the proposed Diffusion Enhancement (DE) network proceeds in the following two processes.

\textit{Forward process.} It transforms the initial data distribution $q(\mathbf{x}_0)$ into a latent variable distribution $q(\mathbf{x}_T)$, where $T$ represents the total number of the timesteps. This transformation follows a fixed Markov chain that can be modeled as:
\begin{equation}\label{eq:qtqt-1}
	q(\mathbf{x}_t|\mathbf{x}_{t-1}) = {\cal N}(\mathbf{x}_t;\sqrt{1-\beta_t}\mathbf{x}_{t-1},\beta_t{\bm I}),
\end{equation}
where $\cal N$, $\{\beta_1,\dots,\beta_T \} \in (1,0)$, and $\bm I$ stand for the Gaussian distribution, a set of hyperparameters and the all-one matrix, respectively. 

By exploiting Eq.~\eqref{eq:qtqt-1}, we have
\begin{equation}
  q(\mathbf{x}_1,\dots,\mathbf{x}_T|\mathbf{x}_0) = \prod_{t=1}^{T}q(\mathbf{x}_t|\mathbf{x}_{t-1}).
\end{equation}

As a result, the forward process can be represented as:
\begin{equation}
  q(\mathbf{x}_t|\mathbf{x}_0) = {\cal N}(\mathbf{x}_t;\sqrt{\bar{a}_t}\mathbf{x}_0,(1-\bar{a}_t){\bm I}),
\end{equation}
where $\alpha_t = 1-\beta_t$ and $\bar{\alpha}_t = \prod_{s=1}^t \alpha_s$. 

Subsequently, we can express $\mathbf{x}_t$ as:
\begin{equation}
  \mathbf{x}_t = \sqrt{\bar{a}_t}\mathbf{x}_0+\sqrt{1-\bar{a}_t}\bm{\epsilon}, 
\end{equation}
where $\bm{\epsilon}\sim {\cal N}(0,{\bm I})$ is a standard Gaussian noise.

\textit{Reverse process.} 
It transforms the latent variable distribution $ p_{\bm \theta}(\mathbf{x}_T)$ back to the data distribution $p_{\bm \theta}(\mathbf{x}_0)$ through a network parameterized by ${\bm \theta}$. 
The reverse process is defined as a Markov chain with learned Gaussian transitions starting with a Gaussian distribution:
\begin{equation}
  p_{\bm \theta}(\mathbf{x}_0, \dots, \mathbf{x}_{T-1}|\mathbf{x}_T) = \prod_{t=1}^{T}p_{\bm \theta}(\mathbf{x}_{t-1}|\mathbf{x}_t),
\end{equation}
where
\begin{equation}
  p_{\bm \theta}(\mathbf{x}_{t-1}|\mathbf{x}_t) = {\cal N}(\mathbf{x}_{t-1};\mu_{\bm \theta}(\mathbf{x}_t,t),\sigma_{\bm \theta} {(\mathbf{x}_t,t)}^2{\bm I}),
\end{equation}
with $\mu_{\bm \theta}(\mathbf{x}_t,t)$ and $\sigma_{\bm \theta} (\mathbf{x}_t,t)$ being the mean and variance of the Gaussian distribution at the $t$-th step.

During the training process, we propose to minimize the mean square error (MSE) loss between the random noise $\bm{\epsilon}$ added to the clean image and the predicted noise $\hat{\bm{\epsilon}}_{\bm \theta}(\mathbf{x}_t,t,\mathbf{y})$ derived from $\mathbf{x}_t$, $t$ and cloudy image $\mathbf{y}$. Since the DE network predicts the noise information based on the cloudy images, it is named conditional noise predictor (CNP). In summary, the loss function employed takes the following form:
\begin{equation} \label{loss}
\mathcal{L}_{{\rm DDPM}} = \mathbb{E}_{\mathbf{x}_t,\bm{\epsilon},t,\mathbf{y}}\Big[{||\bm{\epsilon}-\hat{\bm{\epsilon}}_{\bm \theta}(\mathbf{x}_t,t,\mathbf{y})||}^2\Big].
\end{equation}

\begin{figure}[tbp]
  \centering
   \includegraphics[width=\linewidth]{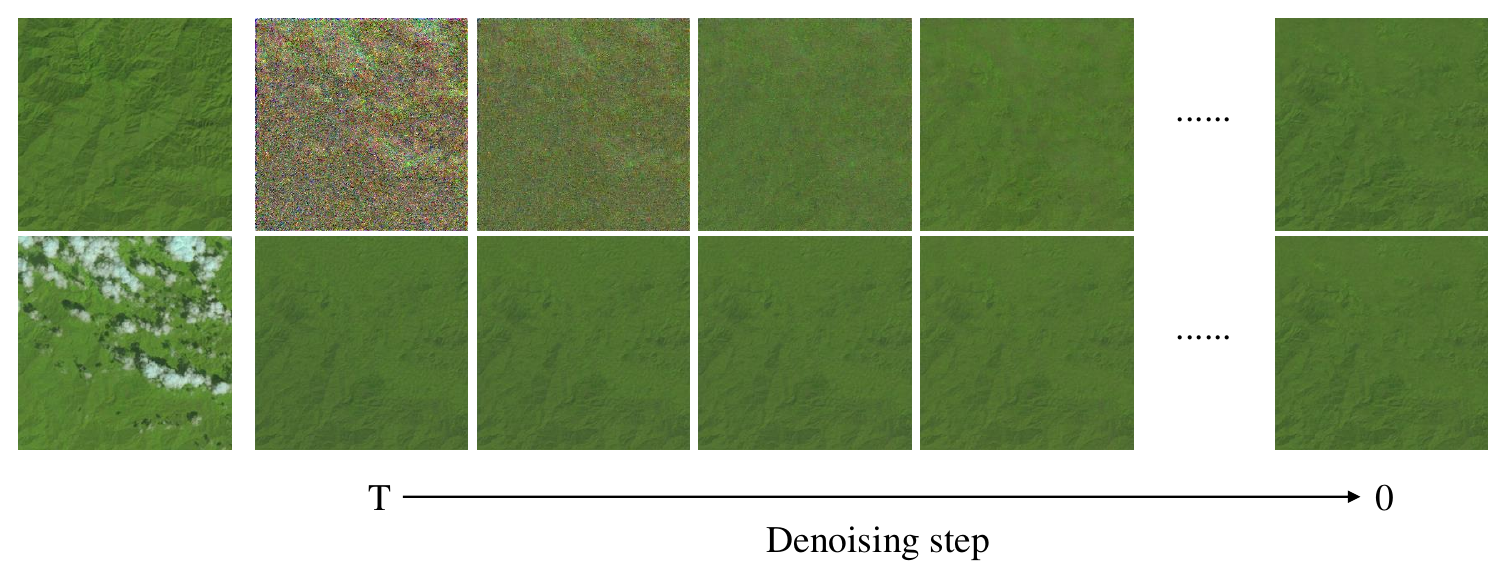}
   \caption{
   The style of $\mathbf{x}_{0, t}$ from denoising time-step $T$ to $0$.
   The first line and second line represent the result of the vanilla diffusion model and our DE, respectively. 
   The cloud-free and cloudy images are presented on the left side.
   }
   \label{fig:denoise}
\end{figure}

\subsection{Reference Visual Prior Integration}

Inspired by \cite{prior1,prior2}, the proposed DE network incorporates an reference visual prior for guiding the inference process to obtain refined results as depicted in Fig.\ref{fig:network}.

For the $t$-th step of the reverse process, $\mathbf{x}_{t-1}$ is calculated based on $\mathbf{x}_{t}$. We first predict the noise $\bm{\epsilon}_t$ based on the state $\mathbf{x}_{t}$ and the time-step $t$.
\begin{equation}
	\bm{\epsilon}_t =\hat{\bm{\epsilon}}_\theta(\mathbf{x}_t, t, \mathbf{y}).
\end{equation}

After that, we obtain $\mathbf{x}_{0, \bm{\epsilon}, t}$ in the current step $t$ based on the predicted noise $\bm{\epsilon}_t$ and the noisy image $\mathbf{x}_t$:
\begin{equation}
	\mathbf{x}_{0, \bm{\epsilon}, t} = (\mathbf{x}_t-\sqrt{1-\bar{\alpha}_t}\bm{\epsilon}_t)/\sqrt{\bar{\alpha}_t}.
	\label{for:x0}
\end{equation}

Through the aforementioned procedure, the formula of one single denoising step is as follows:
\begin{equation}  \label{xt-1}
	\mathbf{x}_{t - 1} =\frac{\sqrt{\bar{\alpha}_{t-1}}\beta_t}{1-\bar{\alpha}_t}\mathbf{x}_{0, t}+\frac{\sqrt{\alpha_t}(1-\bar{\alpha}_{t-1})}{1-\bar{\alpha}_t} \mathbf{x}_{t}+ \tilde{\beta_t}z, z \sim {\cal N}(0,{\bm I}),
\end{equation}
where $\tilde{\beta_t} = \frac{1-\bar{\alpha}_{t-1}}{1-\bar{\alpha}_t}\beta_t$.

In Fig.~\ref{fig:denoise}, we illustrate the example of $\mathbf{x}_{0, t}$ from time-step $T$ to 1 to show the effect of integrating the reference visual prior. As shown in Fig.~\ref{fig:denoise}, with $\mathbf{x}_{0, t} = \mathbf{x}_{0, \bm{\epsilon}, t}$, the diffusion model struggles to remove all the noise and clouds in the initial stages. The quality of $\mathbf{x}_{0, \bm{\epsilon}, t}$ gradually improves from time-step $T$ to $1$. This prolonged iterative process impacts the model's performance, significantly increasing time complexity, and reducing the efficiency in both training and evaluation. However, the textures generated only by the diffusion model often fail to precisely align with the actual scene since the diffusion model primarily focuses on learning the distribution of the entire image set rather than the pixel-level fine information.

Conversely, the reference model implemented in an end-to-end manner primarily rely on fidelity-driven loss functions in their training processes to minimize the pixel differences between cloudy and cloud-free images. As a result, they can reconstruct the underlying structure of cloud-free images to less focus on predicting fine-grained textures with less complexity. This characteristics of reference models make them work well for low-resolution datasets. However, for higher-resolution scenes with richer textures, the reference model fails to capture those fine-grained details. Thus, it is challenging to faithfully restore complicated landscapes concealed beneath the cloud cover.

Considering these pros and cons, we can leverage guidance from an approximately cloud-free image generated by a reference model, denoted as $\mathbf{x}_{0, E}$, to guide the denoising process. $\mathbf{x}_{0, E}$ predicted by the reference model establishes the fundamental image structure, while $\mathbf{x}_{0, \bm{\epsilon}, t}$ generated by the diffusion model introduces realistic details and textures. Furthermore, it is worth noting that reference model results might introduce inaccuracy, and our model integrating the reference visual prior with the progressive diffusion process helps avoid the accumulation of errors. As demonstrated in the second line of Fig.~\ref{fig:denoise}, through this integration, our diffusion enhancement process can bypass the denoising of $\mathbf{x}_0$ and effectively address the limitations of previous reference visual priors.

For the traditional diffusion architecture, $\mathbf{x}_{0, t}$ is equal to $\mathbf{x}_{0, \bm{\epsilon}, t}$. However, in our approach, we utilize an reference visual prior integration for the operation of $\mathbf{x}_{0, t}$ to refine the results of CNP. Specifically, we begin by utilizing the reference model denoted as $\mathbf{E}$ to produce a cloud-free output denoted as $\mathbf{x}_{0, E}$:

\begin{equation}
    \mathbf{x}_{0, E} = \mathbf{E}(\mathbf{y}).
\end{equation}

The output $\mathbf{x}_{0, E}$ generated by the reference model serves as the primary structural foundation of the image, while $\mathbf{x}_{0, \bm{\epsilon}, t}$ predicted by the diffusion model introduces authentic details and textures. A comprehensive formula for this refinement process is presented as follows:
\begin{equation}
    \mathbf{x}_{0, t} = \Gamma(\mathbf{x}_{0, E}, \mathbf{x}_{0, \bm{\epsilon}, t}).
\end{equation}

In practice, we utilize a pixel-wise linear combination of the two predictions:
\begin{equation}  \label{x0}
    \mathbf{x}_{0, t} = (1 - \mathbf{W}) \odot \mathbf{x}_{0, \bm{\epsilon}, t} + \mathbf{W} \odot \mathbf{x}_{0, E},
\end{equation}
where $\odot$ represents the element-wise multiplication and $\mathbf{W} \in\mathbb{R}^{C\times H \times W}$ is a pixel-wise weighting map which will be further described in the next part.

Finally, We apply $\mathbf{x}_{0, t}$ calculated in Eq.~(\ref{x0}) to Eq.~(\ref{xt-1}) to get $\mathbf{x}_{t-1}$ for the denoising step.

\begin{figure}[t]
  \centering
   \includegraphics[width=0.8\linewidth]{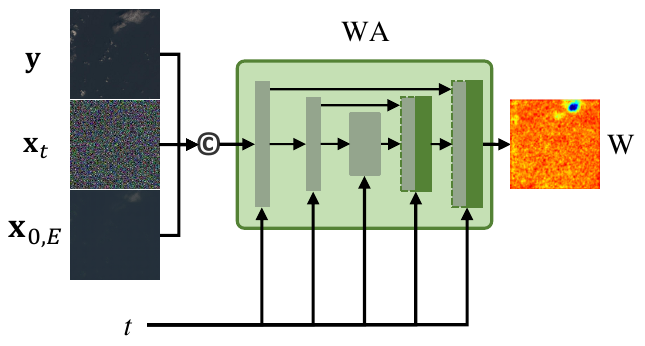}
   \caption{The architecture of Weight Allocation (WA). WA learns to dynamically determine the weighting matrix based on the image features and the noise strength.}
   \label{fig:WA}
\end{figure}

\begin{table*}
	\centering
	\caption{The quantitative experimental results on the RICE1 and RICE2 datasets. $\uparrow$ and $\downarrow$ represent higher better and lower better, respectively.}
	\setlength{\tabcolsep}{4mm}{
		\begin{tabular}{c|ccc|ccc}
			\hline
			\multirow{2}{1cm}{\textbf{Method}}&
			\multicolumn{3}{c|}{\textbf{RICE1}}&\multicolumn{3}{c}{\textbf{RICE2}}\cr
			&\textbf{PSNR}$\uparrow$&\textbf{SSIM}$\uparrow$&\textbf{LPIPS}$\downarrow$&\textbf{PSNR}$\uparrow$&\textbf{SSIM}$\uparrow$&\textbf{LPIPS}$\downarrow$\cr
			\hline
                SpA-GAN   & 28.509 & 0.9122 & 0.0503 & 28.783 & 0.7884 & 0.0963   \\
			AMGAN-CR  & 26.497 & 0.9120 & 0.0447 & 28.336 & 0.7819 & 0.1105   \\
                CVAE      & 31.112 & 0.9733 & 0.0149 & 30.063 & 0.8474 & 0.0743   \\
			MemoryNet & 34.427 & 0.9865 & 0.0067 & 32.889 & 0.8801 & 0.0557   \\
			MSDA-CR   & 34.569 & 0.9871 & 0.0073 & 33.166 & 0.8793 & 0.0463   \\
			\hline
                DE-MemoryNet & \textbf{35.525} & \textbf{0.9882} & \textbf{0.0055} & \textbf{33.637} & 0.8825 & 0.0476  \\
			DE-MSDA   & 35.357 & 0.9878 & 0.0061 & 33.598 & \textbf{0.8842} & \textbf{0.0452}  \\
			\hline
	\end{tabular}}
	\label{tab:rice}
\end{table*}

\begin{table*}
	\centering
	\caption{The quantitative experimental results on the CUHK-CR1 and CUHK-CR2 datasets.$\uparrow$ and $\downarrow$ represent higher better and lower better, respectively.}
	\setlength{\tabcolsep}{4mm}{
		\begin{tabular}{c|ccc|ccc}
			\hline
			\multirow{2}{1cm}{\textbf{Method}}&
			\multicolumn{3}{c|}{\textbf{CUHK-CR1}}&\multicolumn{3}{c}{\textbf{CUHK-CR2}}\cr
			&\textbf{PSNR}$\uparrow$&\textbf{SSIM}$\uparrow$&\textbf{LPIPS}$\downarrow$&\textbf{PSNR}$\uparrow$&\textbf{SSIM}$\uparrow$&\textbf{LPIPS}$\downarrow$\cr
			\hline
                SpA-GAN   & 20.999 & 0.5162 & 0.0830 & 19.680 & 0.3952 & 0.1201   \\
			AMGAN-CR  & 20.867 & 0.4986 & 0.1075 & 20.172 & 0.4900 & 0.0932   \\
                CVAE      & 24.252 & 0.7252 & 0.1075 & 22.631 & 0.6302 & 0.0489  \\
			MemoryNet & 26.073 & 0.7741 & 0.0315 & 24.224 & 0.6838 & 0.0403   \\
			MSDA-CR   & 25.435 & 0.7483 & 0.0374 & 23.755 & 0.6661 & 0.0433 \\
			\hline
                DE-MemoryNet & \textbf{26.183} & \textbf{0.7746} & \textbf{0.0290} & \textbf{24.348} & \textbf{0.6843} & \textbf{0.0369}  \\
                DE-MSDA   & 25.739 & 0.7592 & 0.0321 & 23.968 & 0.6737 & 0.0372  \\
			\hline
	\end{tabular}}
	\label{tab:cuhk}
\end{table*}

\subsection{Dynamic Fusion Among Diffusion Steps}
We employ a Weight Allocation (WA) network, trained to fuse the results of two branches dynamically during steps of the progressive diffusion process. As shown in Fig.~\ref{fig:WA}, the WA takes inputs with the concatenation of $\mathbf{x}_t$, $\mathbf{y}$, and $\mathbf{x}_{0, E}$ while the variable $t$ guides the network across all layers. The UNet architecture of WA is inspired by CNP \cite{guidediff}. Consequently, the training objective allows the WA  dynamically to determine the weighting matrix $\bf W$ based on the image features and the noise strength.

The images produced by the diffusion model initially contain a significant amount of noise, progressively improving as the time-step nears zero. It is essential to adjust the fusion ratio of the results from the reference visual prior. Specifically, the fusion ratio should be consistently effective and gradually diminish as $t$ increases. Furthermore, since image noise is randomly distributed across the entire image and the error from the reference model is uncertain, the fusion factor is also critical. 

To address these challenges, we train the WA network based on the time-step $t$ and the image restoration result of the reference model. It can generate a specific weight $\bf W$ for each time-step $t$, providing detailed pixel-level weight information for the refinement process. Moreover, since $\mathbf{x}_{0, t}$ becomes overwhelmingly dependent on $\mathbf{x}_{0, \bm{\epsilon}, t}$ with the low value of $\bf W$, we introduce a limiting factor $\eta$ to restrain the range of $\bf W$ from $\eta$ to 1 in the inference process. More information about the hyper-parameter $\eta$ is provided in Section~\ref{sec:experiments} -D. In summary, the WA network could accelerate the denoising process and encourage the diffusion model to focus on generating more detailed texture information.

\subsection{Coarse-to-fine Training and Inference}

To accelerate the convergence speed of our Diffusion Enhancement (DE) during the training phase, we implement a coarse-to-fine training strategy. Initially, the image is resized to 1/4 of its original dimensions and processed by a sole diffusion model. Throughout this process, the employed loss function is given in Eq.~(\ref{loss}). The fine-tuning process takes place after the diffusion model reaches near convergence at this smaller scale.

Once the network convergences on the smaller images, we introduce and train the WA network, leveraging the knowledge from the well-converged diffusion network. The WA achieves initial convergence based on the locked diffusion model trained on the downscaled images, laying a foundation for the subsequent joint training of the diffusion model and the WA. In this context, the corresponding loss function of the DE is defined as:

\begin{eqnarray}
  \mathcal{L}_{WA} &=&|\tilde{\mathbf{x}}_0 -\mathbf{x}_{0, t}|\\
 \nonumber &=&|\tilde{\mathbf{x}}_0 - (1 - \mathbf{W}) \odot (\mathbf{x}_{0,\bm{\epsilon}, t})_{sg} + \mathbf{W} \odot \mathbf{x}_{0, E}|,
\end{eqnarray}
where $\tilde{\mathbf{x}}_0$ means the real cloud-free image and $(\cdot)_{sg}$ means stop gradient. Only the gradient of $\bf W$ is calculated while $\mathbf{x}_{0, \bm{\epsilon}, t}$'s is disabled.

Ultimately, the CNP and WA are jointly trained using the full-size image. The loss function for this joint training is defined as:

\begin{equation}
 \mathcal{L}_{joint} = \lambda \cdot \mathcal{L}_{{\rm DDPM}} +  \mathcal{L}_{WA},
\end{equation}
where $\lambda$ is the weight proportion coefficient to balance the value gap between the two parts of the loss function. 
The detailed setting of $\lambda$ is provided in Section \ref{sec:experiments} -B.

Notably, in the second segment of the loss function, the gradient of $\mathbf{x}_{0, \bm{\epsilon}, t}$ remains deactivated to prevent any adverse effects on the CNP. The CNP consistently maintains its original training strategy for larger images, while the WA adapts its approach based on the training outcomes of the CNP.

\begin{figure*}[tbp]
  \centering
   \includegraphics[width=\linewidth]{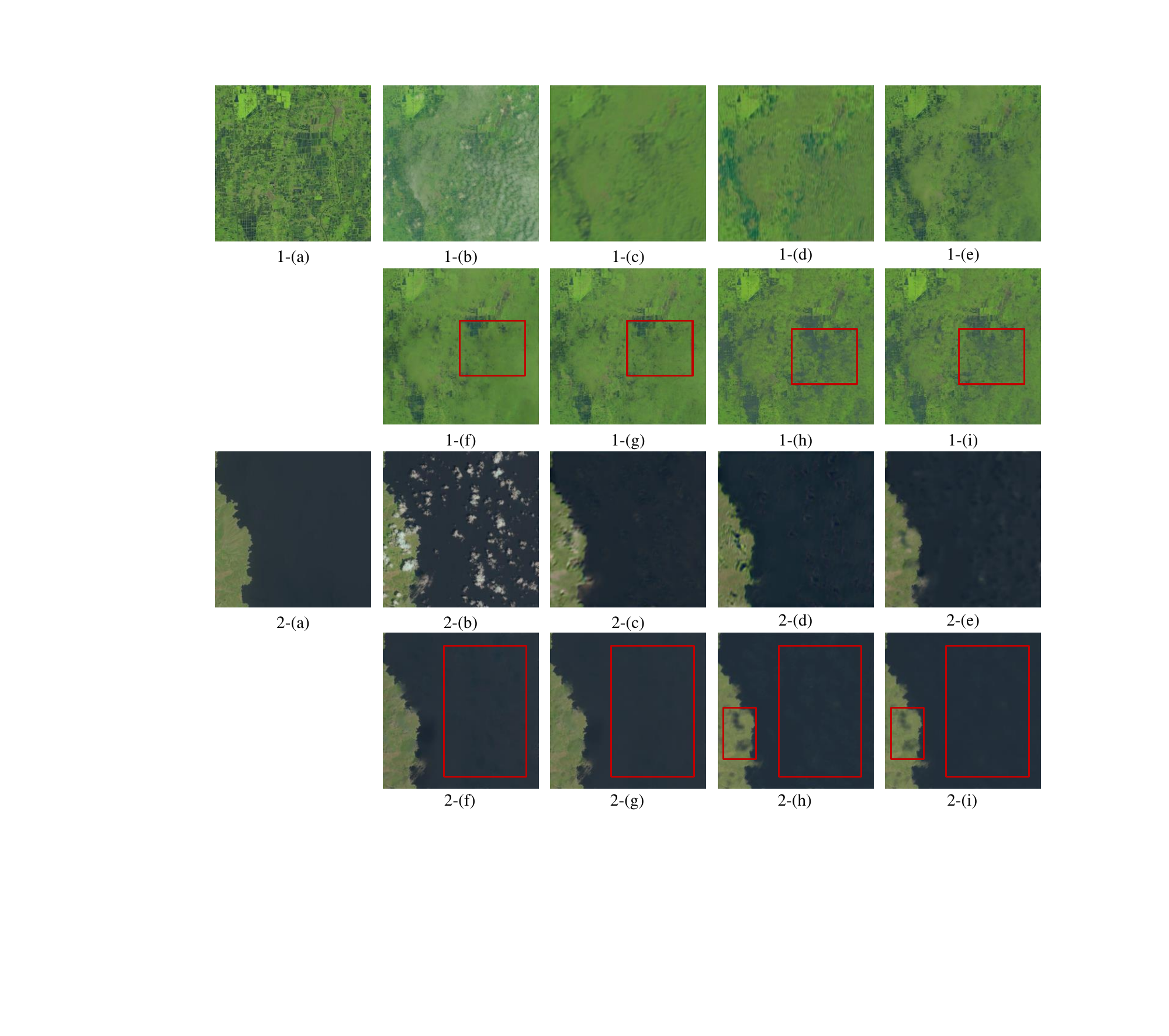}
   \caption{
   Visual comparisons on RICE. (a) Label. (b) Cloudy image. (c) SpAGAN. (d) AMGAN-CR. (e) CVAE. (f) MemoryNet. (g) DE-MemoryNet. (h) MSDA-CR. (i) DE-MSDA.}
   \label{fig:RICE}
\end{figure*}

Throughout the inference process, at each step, the diffusion model predicts the noise $\bm{\epsilon}_t$ and computes $\mathbf{x}_{0, \bm{\epsilon}, t}$ using Eq.~\eqref{for:x0}. Subsequently, the reference model generates its cloud-free output, $\mathbf{x}_{0, E}$, which is then utilized by the WA to determine the weighting map, $\bf W$. $\mathbf{x}_{0, t}$ is calculated through a pixel-wise linear combination of the predictions of $\mathbf{x}_{0, E}$ and $\mathbf{x}_{0, \bm{\epsilon}, t}$ based on the weight $\bf W$ produced by the WA. Ultimately, $\mathbf{x}_{t - 1}$ is generated and the denoising cycle concludes when $t=1$.

\section{Experiments}\label{sec:experiments}
\subsection{Datasets and Metrics}

To evaluate the efficiency of our proposed method, we utilize two datasets: RICE \cite{rice} and the newly introduced CUHK-CR, for validation. The RICE dataset comprises $500$ images with thin cloud covers and 736 images with thick cloud covers in RGB channel and sized at $512 \times 512$ pixels. The training and test sets are randomly partitioned in an 8:2 ratio. Further details about our CUHK-CR dataset are provided in Section~\ref{sec:method}.

We employ three widely recognized metrics for quantitative evaluation of CR performance: Peak Signal-to-Noise Ratio (PSNR), Structural Similarity (SSIM), and Learned Perceptual Image Patch Similarity (LPIPS) \cite{lpips}. 
PSNR evaluates the generated image by comparing it with the ground truth at the pixel level. 
SSIM primarily assesses structural differences, while LPIPS aligns more closely with human perception.

\subsection{Implementation Details}
Our DE is based on the guided diffusion \cite{guidediff}. The hyperparameters of the UNet for CNP and the WA are listed in Table~\ref{tab:model}.

\begin{table}[h]
	\centering
	\caption{The detail model setting of the CNP and WA.}
	\setlength{\tabcolsep}{1.3mm}{
		\begin{tabular}{ccc}
			\hline
                \centering\textbf{Model setting}&\textbf{CNP}&\textbf{WA}\\
                \hline
                Channels&96&64\\
                Depth& 2&2\\
                Channels multiple& 1, 1, 2, 2, 3&1, 1, 2\\
                Attention resolution& 4,8&4,8\\
                Heads&4&1\\
                Dropout& 0.0&0.0\\
		\hline
	\end{tabular}}
	\label{tab:model}
\end{table}

\begin{figure*}[tbp]
  \centering
   \includegraphics[width=\linewidth]{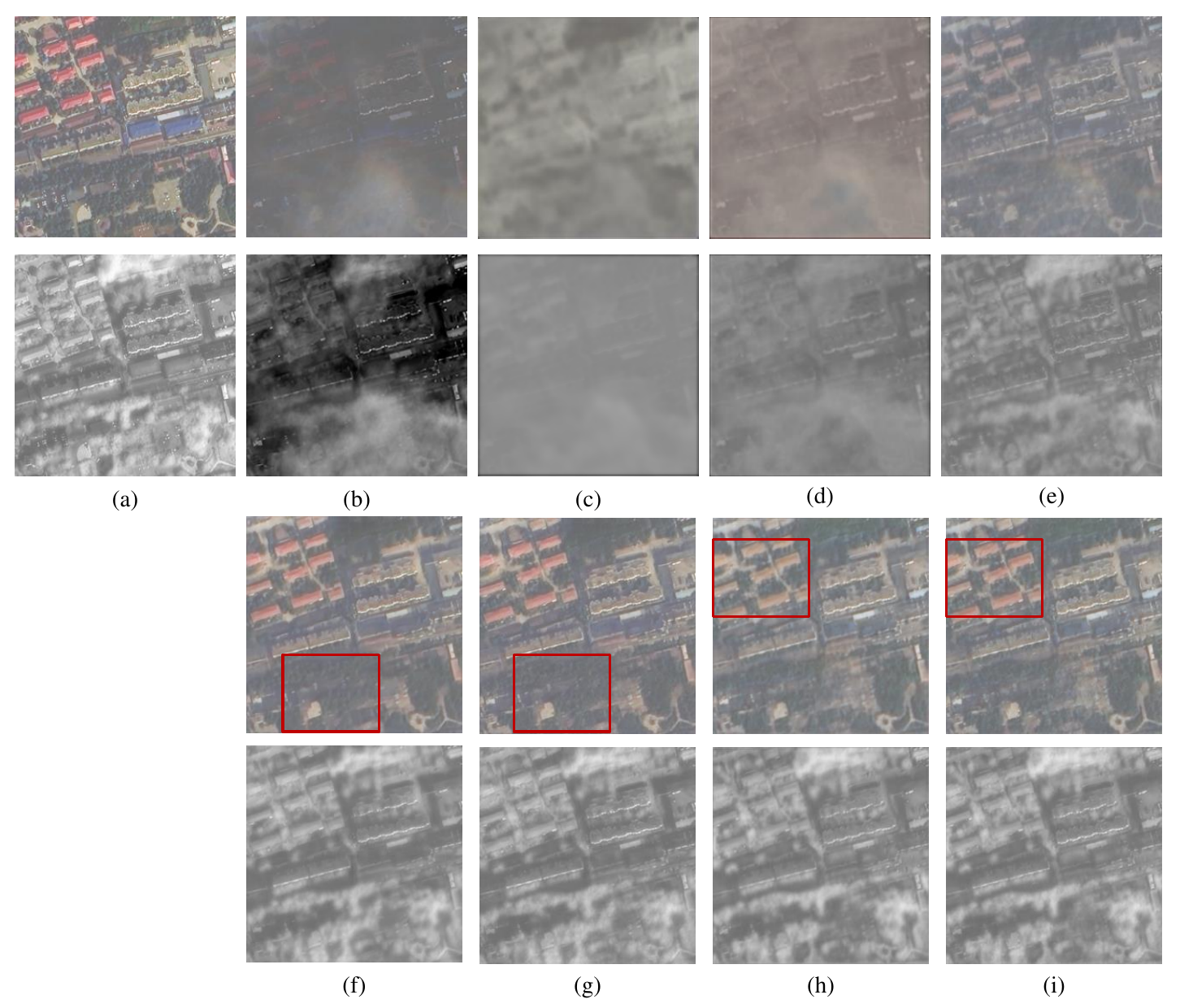}
   \caption{Visual comparisons on CUHK-CR. The first line and second line present the RGB images and near-infrared images, respectively. (a) Label. (b) Cloudy image. (c) SpAGAN. (d) AMGAN-CR. (e) CVAE. (f) MemoryNet. (g) DE-MemoryNet. (h) MSDA-CR. (i) DE-MSDA.}
   \label{fig:CUHK-CR}
\end{figure*}

In DE, the CNP and WA undergo training employing the L2 and L1 loss, respectively, with a consistent learning rate of $10^{-5}$. We maintain a weight proportion coefficient, $\lambda$, set at 1. To enhance efficiency in inference, we implement DDIM \cite{ddim} with 50 steps, and the limiting factor $\eta$ is set to 0.3 which means that the values of $\bf W$ are confined within the range of 0.3 to 1. All images, for both training and testing, are standardized to dimensions of 256 $\times$ 256 pixels. Initially, CNP is trained by smaller images measuring 64 $\times$ 64 pixels, utilizing a batch size of 64. As the training dataset shifts to standard-sized 256 $\times$ 256 pixel images, the batch size is adjusted to 16. For our CUHK-CR dataset, we perform model training and testing using 4-band multispectral images. All experiments are executed using PyTorch on a single NVIDIA GeForce RTX 4090 GPU equipped with 24 GB of RAM.

\subsection{Performance Comparison}

We conduct a comprehensive comparison between our DE and several state-of-the-art CR networks, including MemoryNet \cite{mn}, CVAE \cite{cvae}, SpA-GAN \cite{spagan}, AMGAN-CR \cite{amgan}, and MSDA-CR \cite{mdsa}. To ensure a fair evaluation, all of these methods are thoroughly optimized using our training and test datasets to achieve their peak performance.

The quantitative results of these experiments on the RICE and CUHK-CR datasets are presented in Table~\ref{tab:rice} and Table~\ref{tab:cuhk}, respectively. Since the visual differences of the thin clouds are not readily discernible, we have chosen to display visual comparisons solely for thick cloud datasets at Fig.~\ref{fig:RICE} and Fig.~\ref{fig:CUHK-CR}.

\subsubsection{RICE}

As indicated in Table~\ref{tab:rice}, our method demonstrates a substantial improvement compared to its corresponding reference model. Notably, we focus our DE on MDSA-CR and MemoryNet, given their superior performance among these end-to-end models. For MSDA, which achieves the best results on both RICE datasets, our DE-MSDA exhibits an improvement of 0.8 dB and 0.001, 0.4 dB, and 0.01 in PSNR and LPIPS for RICE1 and RICE2, respectively. These gains in LPIPS indicate that our results align with human perception better. Our diffusion-based approach significantly enhances the generation of fine textures, closely matching the ground truth, within the framework provided by the corresponding reference model. The enhancements on LPIPS are especially obvious in the context of RICE2, where the dense cloud cover raises a hard challenge for cloud-free image reconstruction. This scenario demands a heightened capacity for generating complicated and visually authentic textural details, given the expanse and a considerable amount of obscured texture concealed by the clouds. Consequently, the model's capability to predict and generate textures is highlighted in such conditions. Though the end-to-end models such as MemoryNet and MSDA-CR also gain promising results, our DE could make additional improvements based on them.

Visual results are presented in Fig.~\ref{fig:RICE}. SpAGAN and AMGAN-CR exhibit obvious shortcomings in image style and color. Although MSDA-CR and MemoryNet yield superior results, there are still some errors, such as residual noise and cloud cover, as indicated by red boxes. Our DE is capable of correcting errors and making detailed predictions. For example, our model effectively removes these artifacts, and in the central part of the island, where blue dominates, our model accurately replaces it with green, making the cavity less conspicuous.

\subsubsection{CUHK-CR}

The restored results of our CUHK-CR dataset are generally less satisfactory compared to RICE. The highest PSNR achieved by the end-to-end model in the RICE dataset surpasses 30 dB, but it decreases to 26 dB and 24 dB in the CUHK-CR1 and CUHK-CR2 datasets, respectively. The results signify that our ultra-resolution dataset presents greater challenges. Despite the increased difficulty, our DE applied to MSDA still yields superior results, achieving nearly a 0.3 dB PSNR improvement in both CUHK-CR1 and CUHK-CR2. On the CUHK-CR dataset, the limitations of certain models like SpAGAN and AMGAN-CR become more conspicuous when confronted with such ultra-resolution images, underscoring their unsuitability for high-resolution CR tasks in the realm of RS. They exhibit limited effectiveness in removing clouds, with an improvement of less than 1 dB improvement over the cloudy image.

Visual results for CUHK-CR are provided in Fig.~\ref{fig:CUHK-CR}. SpAGAN and AMGAN-CR struggle with such high-resolution CR tasks, particularly in the presence of thick clouds. In the case of CVAE, despite a reasonable outline, it grapples with severe color deviations. Our DE primarily introduces subtle changes and correcting color errors when compared to their corresponding reference models. For instance, the color of the roof in DE-MSDA's output more closely resembles the ground truth than that of MSDA-CR. Furthermore, the results after our enhancement appear clearer and more accurate, in contrast to the rather blurry outputs of MemoryNet, especially in areas obscured by clouds.

\subsection{WA Analysis}

\subsubsection{Spatial adaptation}

\begin{figure}[t]
  \centering
   \includegraphics[width=\linewidth]{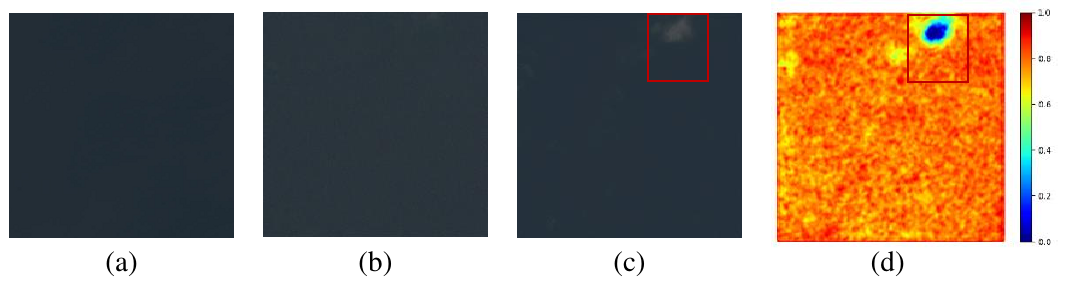}
   \caption{A example presents the attention heatmap of $\bf W$. As the value approaches 1, its reliance on $\mathbf{x}_{0, E}$ becomes more pronounced. Conversely, as it nears 0, it exhibits a stronger dependence on $\mathbf{x}_{0, \bm{\epsilon}, t}$. (a) Label. (b) $\mathbf{x}_{0, \bm{\epsilon}, t}$. (c) $\mathbf{x}_{0, E}$. (d) Heatmap of $\bf W$. }
   \label{fig:hotmap}
\end{figure}

In Fig.~\ref{fig:hotmap}, we present an example of the attention heat map that depicts the behavior of the WA. Notably, the reference model falls short of completely removing the cloud cover, as indicated by the highlighted area within the red box. As depicted in Fig.~\ref{fig:hotmap} (d), our WA diligently addresses this discrepancy by reducing the weight allocated to this specific area. Moving to the domain of $\mathbf{x}_{0, \bm{\epsilon}, t}$, we observe that some regions still retain residual noise that has not been eliminated. In response, the value of $\bf W$ is notably higher in these challenging areas, denoting slight adjustments that correspond to the noise distribution. This attention heat map serves as a compelling visual representation of the WA's capacity to dynamically fine-tune the strength of the reference visual prior integration. The results demonstrate that this fine-tuning process is based on the assessment of the quality of both $\mathbf{x}_{0, E}$ and $\mathbf{x}_{0, \bm{\epsilon}, t}$, ensuring that the CR process is optimized for various image regions.

\begin{figure}[t]
  \centering
   \includegraphics[width=0.8\linewidth]{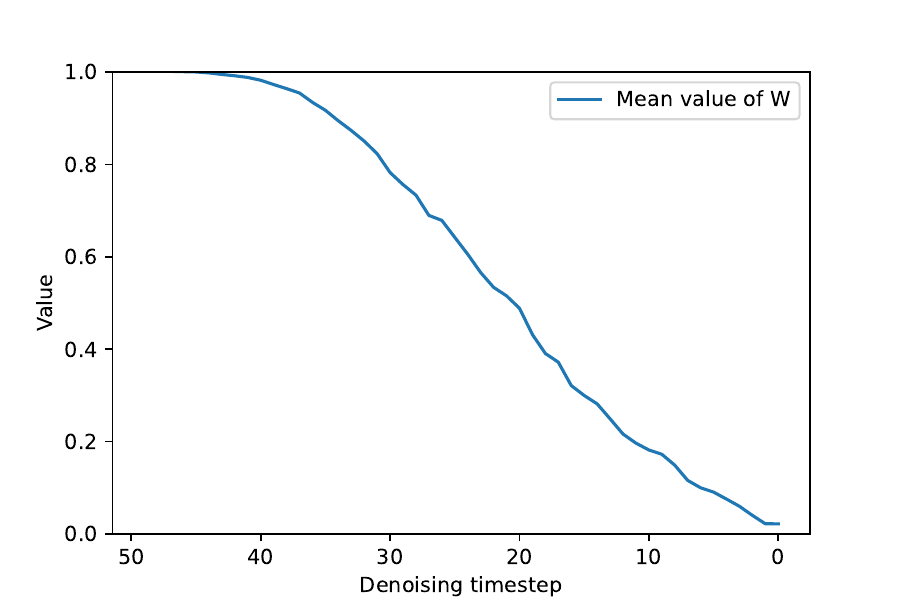}
   \caption{The mean value generated by the WA in each time-step on RICE2 with reference model MSDA-CR. }
   \label{fig:trend}
\end{figure}

\subsubsection{Dynamic fusion among different diffusion steps}

The change of the mean value of $\bf W$ over each time-step is visually represented in Fig.~\ref{fig:trend}. 
Initially, the mean value of $\bf W$ is relatively high and gradually decreases as the time-step decreases, 
approaching nearly 0 in the later stages. 
This trend indicates that, at the outset, $\mathbf{x}_{0, t}$ primarily relies on the guidance provided by $\mathbf{x}_{0, E}$, while the influence of $\mathbf{x}_{0, \bm{\epsilon}, t}$ becomes more prominent as the time-step approaches 0.

The fluctuation in the mean value of $\bf W$ reveals the underlying assumption that prior integration establishes the groundwork for the overall structure of $\mathbf{x}_{0, t}$ in the first few denoising steps, outlining the likely shapes of the images. Subsequently, the diffusion architecture intervenes, making fine adjustments by introducing additional texture information and correcting errors based on the guidance. This dynamic shift in the mean value of $\bf W$ underscores the collaborative relationship between the prior integration and the diffusion architecture, leading to a leap in reconstruction performance.

\subsubsection{Parameter Analysis}

\begin{figure}[t]
  \centering
   \includegraphics[width=\linewidth]{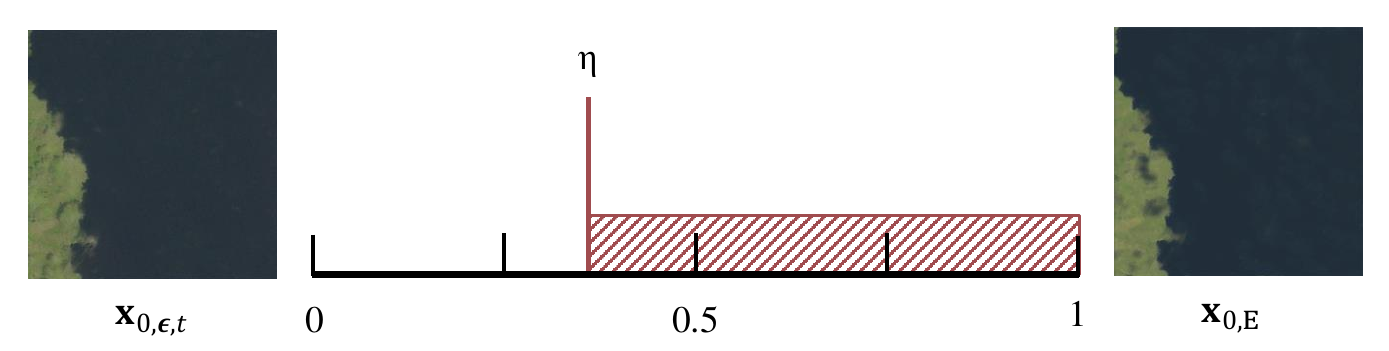}
   \caption{The schematic representation of the adjustment of the limiting factor $\eta$. The red box means the limited value range of $\bf W$. }
   \label{fig:exampeWlimit}
\end{figure}

\begin{figure}[t]
  \centering
   \includegraphics[width=0.8\linewidth]{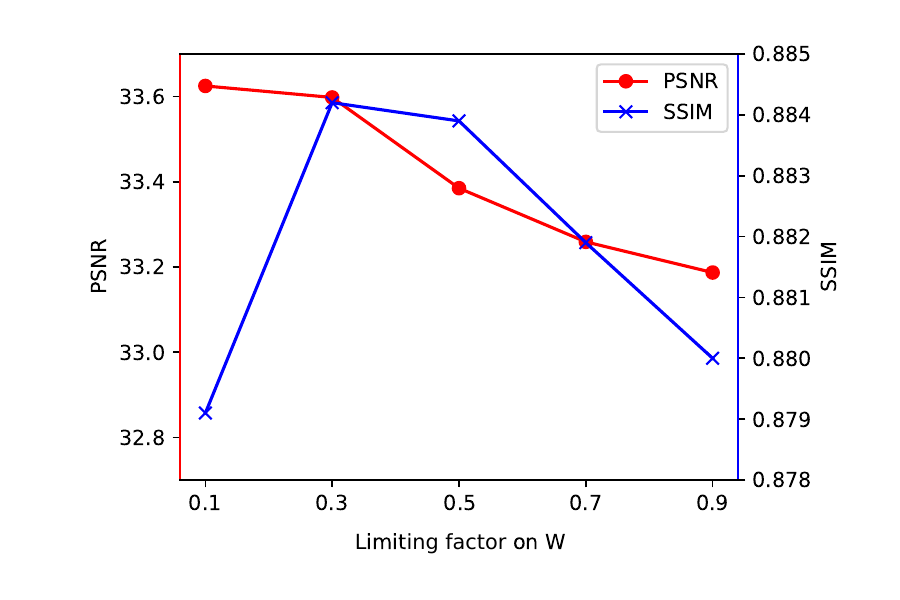}
   \caption{Experimental comparisons on RICE with different limiting factor $\eta$ on $\bf W$. }
   \label{fig:limitW}
\end{figure}

Our investigation delves into the impact of the limiting factor $\eta$ on $\bf W$. The schematic representation of $\eta$ adjustment is thoughtfully illustrated in Fig.~\ref{fig:exampeWlimit}. It means that $\bf W$ is limited at the range of $\eta$ to 1. In the training process, we set $\eta$ to 0, thereby effectively allowing $\bf W$ to range from 0 to 1 without any constraints. The WA network flexibly learns the balance between $\mathbf{x}_{0, \bm{\epsilon}, t}$ and $\mathbf{x}_{0, E}$. However, in the inference process, with low values of $\bf W$, $\mathbf{x}_{0, t}$ becomes overwhelmingly dependent on $\mathbf{x}_{0, \bm{\epsilon}, t}$. In this scenario, $\mathbf{x}_{0, t}$ may incorporate a substantial amount of inaccurate information. To address this concern, we set the limiting factor $\eta$ to a value more than 0 to restrict the range of values for $\bf W$. In theory, $\eta$ serves to control the maximum influence that $\mathbf{x}_{0, \bm{\epsilon}, t}$ can exert based on the reference visual prior refinement. Our evaluation of various $\eta$ values, including $\{0.1, 0.3, 0.5, 0.7, 0.9\}$, reveal interesting insights. We note that our DE achieves the highest PSNR when $\eta$ is set to 0.1, while the SSIM is maximized when $\eta$ is set to 0.3, as illustrated in Fig.~\ref{fig:limitW}. In summary, our DE appears to yield the most favorable results when $\eta$ is set to 0.3, achieving the performance balance between structural detail and global contour preservation. This optimized setting of $\eta$ ensures that both $\mathbf{x}_{0, E}$ and $\mathbf{x}_{0, \bm{\epsilon}, t}$ contribute effectively to the cloud-free image generation process.

\subsection{Gap between the high-resolution and low-resolution datasets}


\begin{table}[t]
	\centering
	\caption{The result of MemoryNet trained with the different resolution images. Training and Testing represent the spatial resolution of images for training and testing.}
	\setlength{\tabcolsep}{1.3mm}{
		\begin{tabular}{ccccc}
			\hline
                \centering\textbf{Training}&\textbf{Testing}&\textbf{PSNR} &\textbf{SSIM}&\textbf{LPIPS} \\
                \hline
                2m& 1m   & 22.773 & 0.6238 & 0.0438\\
                2m& 0.5m & 21.393 & 0.5934 & 0.0529\\
			1m& 1m   & 24.819 & 0.7098 & 0.0392\\
                1m& 0.5m & 22.575 & 0.6503 & 0.0443\\
		\hline
	\end{tabular}}
	\label{tab:image_size}
\end{table}

We perform extra experiments to demonstrate the significant influence of differences in data resolution on the model's performance. Essentially, models trained with low-resolution images yield less favorable results when tested on high-resolution datasets. This emphasizes the necessity for an ultra-resolution CR dataset.

Our approach initiates by training the model with images of various resolutions in the same size, followed by evaluating its performance on high-resolution sets. Specifically, we resize our 512 $\times$ 512 images from 0.5m to different spatial resolution, such as $\{1m, 2m\}$, and crop all of them as 128 $\times$ 128 to train the model. Following the training phase, we utilize corresponding crop sizes 128 $\times$ 128 from the original and resized images with 0.5m and 1m spatial resolution to assess the impact of resolution on the ultimate CR results. As depicted in Table~\ref{tab:image_size}, all metrics show degradation as the training image resolution decreases. When comparing the performance between the training spatial resolution of 1m and 2m on 0.5m test set, we observe a decrease of 1.2 dB in PSNR, 0.06 in SSIM, and 0.008 in LPIPS. These experimental findings underscore the importance of our efforts to construct an ultra-resolution CUHK-CR dataset.

\subsection{Ablation Study}

\begin{table}[t]
	\centering
	\caption{Vertical ablation study on the RICE2 with MSDA-CR.}
	\setlength{\tabcolsep}{1.3mm}{
		\begin{tabular}{ccccccc}
			\hline		
                \centering\textbf{No.}&\textbf{Image size}&\textbf{Diffusion Fix}&\textbf{WA}& \textbf{PSNR} &\textbf{SSIM}&\textbf{LPIPS} \\
                \hline
                1&64&No&& 31.408 & 0.8454 & 0.0518\\
                2&256&Yes&\checkmark& 33.573 & 0.8792 & 0.0465\\
			3&256&No&\checkmark& 33.598 & 0.8842 & 0.0452\\
		\hline
	\end{tabular}}
	\label{tab:Ablation}
\end{table}

Table \ref{tab:Ablation} illustrates the results of an ablation study that explores the impact of the coarse-to-fine training strategy, WA, and reference visual prior integration. The results are presented in the order of training steps. No. 1 represents the outcome of training the diffusion model solely with small images, while No. 2 denotes the results based on the pre-trained model from No. 1, which is further trained with regular-sized images on WA. No. 3 is the final result, where the WA and the diffusion model are jointly fine-tuned with normal-sized images. In comparison to No. 1 and No. 2, the inclusion of WA and reference visual prior refinement results in a remarkable improvement of nearly 2.1 dB, 0.034, and 0.005 in terms of PSNR, SSIM, and LPIPS, respectively. The fine-tuning process on normal-sized images has a lesser impact on PSNR and LPIPS but contributes more significantly to SSIM with a 0.005 improvement. These experimental results emphasize the advantageous role of coarse-to-fine training strategy, WA, and reference visual prior integration in the training order.

\begin{table}[t]
	\centering
	\caption{Horizontal ablation study on the RICE2 with MSDA-CR.}
	\setlength{\tabcolsep}{1.3mm}{
		\begin{tabular}{ccccc}
			\hline		
                \centering\textbf{Coarse-to-fine}&\textbf{WA}& \textbf{PSNR} &\textbf{SSIM}&\textbf{LPIPS} \\
                \hline
                \checkmark&& 33.215 & 0.8835 & 0.0457\\
                &\checkmark& 33.297 & 0.8820 & 0.0444\\
			\checkmark&\checkmark& 33.598 & 0.8842 & 0.0452\\
		\hline
	\end{tabular}}
	\label{tab:Ablation2}
\end{table}

In the previous paragraph, we illustrated the improvements achieved through our three-stage experimental process. Here, we conduct a horizontal comparison by presenting results without the coarse-to-fine training strategy and the WA in Table \ref{tab:Ablation2}. The first row shows the outcomes from models trained solely with normal-sized images, excluding the coarse-to-fine training strategy, while the second row showcases results in which the WA is replaced with a simple linear combination using a fixed parameter of 0.5. In other words, both $\mathbf{x}_{0, \bm{\epsilon}, t}$ and $\mathbf{x}_{0,\mathbf{E}}$ each contribute half to $\mathbf{x}_{0,t}$ at any time-step. As indicated in Table \ref{tab:Ablation2}, the inclusion of the WA results in an improvement of nearly 0.4 dB in PSNR, while the coarse-to-fine training strategy leads to a gain of 0.3 dB in PSNR, 0.002 in SSIM, and 0.001 in LPIPS. This horizontal comparison objectively highlights the advantages of the WA and the coarse-to-fine training strategy.

\subsection{Computational Complexity Analysis}

\begin{table}[t]
	\centering
	\caption{Computational complexity of the compared methods.}
	\setlength{\tabcolsep}{1.5mm}{
		\begin{tabular}{cccccc}
			\hline
			\multirow{2}{1cm}{\textbf{Models}}&\textbf{Complexity}&\textbf{Memory}&\textbf{Parameters}&\textbf{Speed}\\
			&\textbf{(G)}&\textbf{(MB)}&\textbf{(M)}&\textbf{(FPS)} \\
			\hline
			SpAGAN&33.97& 6196&0.22&29.37\\
                AMGAN-CR&96.97&5668&0.24&28.94\\
			CVAE&92.88&6068&15.56&37.73\\
			MN&1097.30&11822&3.64&17.61\\
			MSDA&106.89&11318&3.91&17.80\\
			\hline
                DE&397.65&13112&36.80& 12.04\\
			\hline
	\end{tabular}}
	\label{tab:complex}
\end{table}

We conduct a thorough comparison of the computational complexity among the models, in terms of model complexity, memory usage, parameter count, and processing speed. The specific details are presented in Table \ref{tab:complex}. The results demonstrate that our model achieves superior results without significantly increasing computational complexity.

\section{Conclusion} \label{sec:conclusion}

In this article, the Diffusion Enhancement (DE) method is introduced for reconstructing cloud-free images. DE incorporates the diffusion architecture under the basis guidance of an reference visual prior integration, aiming to capture the merits of progressive diffusion process and convolutional neural network to achieve both excellent global context modeling and fine-grained detailed reconstruction. To adaptively fuse the information from both branches, a Weight Allocation (WA) network is trained to make balance based on their outputs across the whole denoising steps. Additionally, a coarse-to-fine training strategy is employed to accelerate convergence while obtaining superior results within a limited number of iterations. Finally, we introduce an ultra-resolution benchmark that provides a new basis with well-defined spatial landscape textures to train and evaluate the performance of CR models. Our experimental results on our DE on both the RICE and our CUHK-CR datasets demonstrate its superior performance. For future works, various conditions, such as feature maps of cloudy images and semantic information, may substitute the cloudy image to offer improved guidance for constructing more effective diffusion models.

\small
\bibliographystyle{IEEEtranN}
\bibliography{references}

\end{document}